# Chaotic Dirac billiard in graphene quantum dots

L. A. Ponomarenko[1], F. Schedin[1], M. I. Katsnelson[2], R. Yang[1], E. H. Hill[1], K. S. Novoselov[1], A. K.Geim[1]
[1]Centre for Mesoscience and Nanotechnology, University of Manchester, Manchester M13 9PL, UK
[2]Institute for Molecules and Materials, University Nijmegen, 6525 ED Nijmegen, The Netherlands

*We report on transport characteristics of quantum dot devices etched entirely in graphene. At large sizes, they behave as conventional single-electron transistors, exhibiting periodic Coulomb blockade peaks. For quantum dots smaller than 100 nm, the peaks become strongly non-periodic indicating a major contribution of quantum confinement. Random peak spacing and its statistics are well described by the theory of chaotic neutrino (Dirac) billiards. Short constrictions of only a few nm in width remain conductive and reveal a confinement gap of up to 0.5eV, which demonstrates the in-principle possibility of molecular-scale electronics based on graphene.*

The exceptional electronic properties of graphene and its formidable potential in various applications have ensured a rapid growth of interest in this new material [1,2]. One of the most discussed and tantalizing directions in research on graphene is its use as the base material for electronic circuitry that is envisaged to consist of nanometer-sized elements. Most attention has so far been focused on graphene nanoribbons (see [3-9] and references therein).

In this Letter, we report quantum dot (QD) devices made entirely from graphene, including their central islands (CI), quantum barriers, source and drain contacts and side-gate electrodes. We have found three basic operational regimes for such devices, depending on their size. For relatively large (submicron) CIs, size quantization plays an insignificant role, and our devices were found to operate as orthodox single-electron transistors (SET) exhibiting periodic Coulomb blockade (CB) oscillations. The CB regime has been extensively studied previously using metallic and semiconducting materials [10,11] and, more recently, the first SET devices made from graphite [12] and graphene [1,13,14] were also demonstrated. The all-graphene SETs reported here are technologically simple, reliable and robust and can operate above liquid-helium temperatures $T$, which makes them attractive candidates for use in various charge-detector schemes [10]. For intermediate CI sizes (less than ~100nm), we enter into the quantum regime, in which the confinement energy $\delta E$ >10meV exceeds the charging energy $E_c$. Such a strong quantization for relatively modest confinement is unique to massless fermions [1,2] and related to the fact that their typical level spacing $\delta E \approx v_F h/2D$ in a quantum box of size $D$ is much larger than the corresponding energy scale $\approx h^2/8mD^2$ for massive carriers in other materials ($v_F \approx 10^6$m/s is the Fermi velocity in graphene, $h$ the Planck constant and $m$ the effective mass). This means that level splitting in graphene-based 100-nm devices should be tens and hundreds times larger than in typical semiconducting and metal QDs, respectively. This regime is probably most interesting from the fundamental physics point of view, allowing studies of relativistic-like quantum effects in confined geometries [15-21]. In particular, we have observed a strong level repulsion in QDs, which is a clear signature of quantum chaos (so-called "neutrino billiards" [15]). Conductance of our smallest devices is dominated by individual constrictions with sizes down to ~1nm, which exhibit $\delta E$ ~0.5eV and a good-quality transistor action at room $T$. It is remarkable that these molecular-scale structures survive microfabrication procedures, remain mechanically and chemically stable and highly conductive under ambient conditions and sustain large (nA) currents.

Our devices were made from graphene crystallites prepared by micromechanical cleavage on top of an oxidized Si wafer (300nm of $SiO_2$) [22]. By using high-resolution electron-beam lithography, we defined a 30-nm thick polymethylmethacrylate (PMMA) mask that protected chosen areas during plasma etching in oxygen and allowed us to carve graphene into a desired geometry. The inset in Fig. 1 shows one of our working devices that generally consisted of the CI of diameter $D$, connected via two short constrictions to wide source and drain regions; the devices also had side-gate electrodes (we often placed them ~1μm away from the CI as explained in Supplementary Information (SI)). The Si wafer was used as a back gate. The constrictions were designed to have equal length and width of 20nm [SI], and we refer to them as quantum point contacts (QPCs). They provided quantum barriers to decouple the CI from the contacts [10,11]. If necessary, by using further plasma etching (after the devices were tested), we could narrow QPCs by several nm, exploiting the fact that the PMMA mask was gradually etched away not only from top but also sideways. This allowed us to tune the resistance of QDs to a value of several hundred kΩ, i.e. significantly larger than resistance quantum $h/e^2$, which is an essential requirement for single electron transport [10,11]. Graphene QDs as small as 30nm in diameter $D$ could be fabricated reliably using this approach (inset in Fig. 1). For even smaller $D$, irregularities in PMMA (~5 nm [23]) became comparable in size with the designed features and, unavoidably on this scale, we could only estimate the device geometry. The measurements discussed below were carried out by using the standard lock-in technique with dc bias over a $T$ range from 0.3 to 300K. To control CB, we used both side and back gates with the latter allowing extensive changes in the population of QD levels whereas the former was useful for accurate sweeps over small energy intervals [SI]. The response to the side-gate potential differed for different devices but could be related to back-gate voltage $V_g$ through a numerical factor. For consistency, all the data below are presented as a function of $V_g$.

Let us start with the behavior observed for our relatively large devices (Fig. 1). They exhibit (nearly) periodic CB resonances that at low $T$ are separated by regions of zero conductance $G$. As $T$ increases, the peaks become broader and overlap, gradually transforming into CB oscillations. The oscillations become weaker, as $G$ increases with carrier concentration or $T$, and completely disappear for $G$ larger than ~$0.5e^2/h$ because the barriers become too transparent to allow CB. For the data in Fig. 1b, we can identify more than 1000 oscillations. Their periodicity, $\Delta V_g \approx 16$mV, yields the capacitance between the back gate and CI, $C_g = e/\Delta V_g \approx 10$aF. For a disk placed on top of SiO$_2$ (dielectric constant $\varepsilon \approx 4$) at a distance $h \geq D$ from the Si gate, the gate capacitance is nearly the same as for an isolated disk [25], that is, $C_g \approx 2\varepsilon_0(\varepsilon+1)D \approx 20$aF for $D = 250$nm. The difference by a factor of 2 can be accounted for in terms of screening by the contact regions [24].

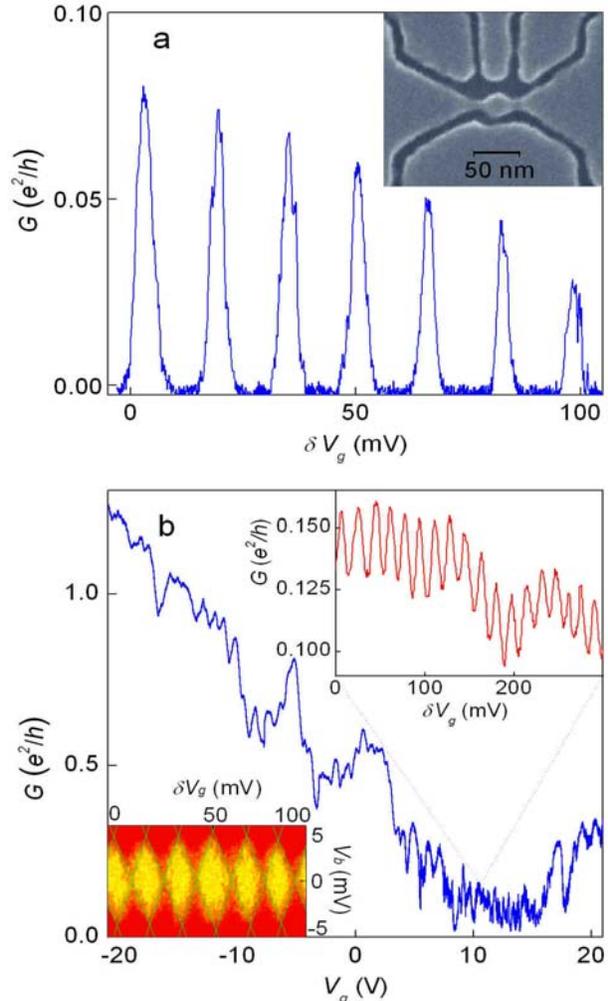

Figure 1. Graphene-based single-electron transistor. **a**, Conductance $G$ of a device with the central island of 250nm in diameter and distant side gates [SI] as function of $V_g$ in the vicinity of +15V; $T = 0.3$K. The inset shows one of our smaller devices to illustrate the high resolution of our electron-beam lithography that allows features down to 10 nm. Dark areas in the scanning electron micrograph are gaps in the PMMA mask so that graphene is removed from these areas by plasma etching. In this case, a 30-nm QD is connected to contact regions through narrow constrictions and there are 4 side gates. **b**, Conductance of the same device as in Fig. 1a over a wide range of $V_g$ ($T = 4$K). Upper inset: Zooming into the low-$G$ region reveals hundreds of CB oscillations. The lower inset shows Coulomb diamonds: differential conductance $G_{diff} = dI/dV$ as a function of $V_g$ (around +10V) and bias $V_b$ (yellow-to-red scale corresponds to $G_{diff}$ varying from zero to $0.3e^2/h$; note that our color diagrams often appear smudged if printed in gray).

The overall shape of the conductance curve $G(V_g)$ in Fig. 1b resembles that of bulk graphene [1] but is distorted by smooth (on the scale of $\Delta V_g$) fluctuations that are typical for mesoscopic devices and are due to quantum interference effects [1-4,13,14]. Smooth variations in the CB peak height (Fig. 1a) are also attributed to interference-induced changes in the barriers' transparency, as shown by studying individual QPCs [SI]. We have also measured the dependence of CB peaks on applied bias $V_b$ and, from the standard stability diagrams (Coulomb diamonds), found the charging energy $E_c$. The lower inset in Fig. 1b shows such diamonds for $D \approx 250$nm, which yields $E_c \approx 3$meV and the total capacitance $C = e^2/E_c \approx 50$aF. Rather large $E_c$ implies that the CB oscillations in Fig. 1b are smeared mostly due to an increase in the barrier transparency with $T$, and submicron graphene SETs can be operational at $T \geq 10$K. In general, the observed CB behavior in our large-$D$ devices is in agreement with the one exhibited by conventional SETs [10,11].

For devices smaller than ~100nm, we observed a qualitative change in behavior: CB peaks were no longer a periodic function of $V_g$ but varied strongly in their spacing. Figs. 2a,b illustrate this behavior for $D \approx 40$nm, whereas Fig. 2c plots the distance $\Delta V_g$ between the nearest peaks for 140 of them. One can see that $\Delta V_g$ varies by a factor of 5 or more, which exceeds by orders of magnitude typical variations of $\Delta V_g$ observed in non-graphene QDs [11,25,26]. This is a clear indication that the size quantization becomes an important factor. Indeed, the distance between CB peaks is determined by the sum of charging and confinement energies $\Delta E = E_c + \delta E$ and the latter contribution is expected to be exceptionally large for massless quasiparticles ($\delta E \approx \alpha/D$ with theory estimates for $\alpha$ ranging from 0.2 to 1.5 eV nm [3-8]). Although the observed large confinement effects are in agreement with theory, we have also considered other mechanisms [SI]. For example, in stochastic CB [27], peaks could become seemingly non-periodic due to the presence of two or more QDs. In our case, however, variations in the peak height were modest, no new peaks appeared with changing $T$, variations in $\Delta V_g$ were one-to-one related to changes in height $\Delta E$ of the Coulomb diamonds, and the same values of $\Delta E$ were found from the $T$ dependence of $G$ between peaks, all in stark contrast to the case of stochastic CB [SI,27]. Typical changes in $\Delta E$ yield characteristic $\delta E$ and the found values agree well with the confinement gaps expected theoretically (for example, $\delta E$ is ~10meV for the 40-nm QD in Fig. 2). Accordingly, we refer to our devices with $D$ <100 nm as QDs rather than SETs.

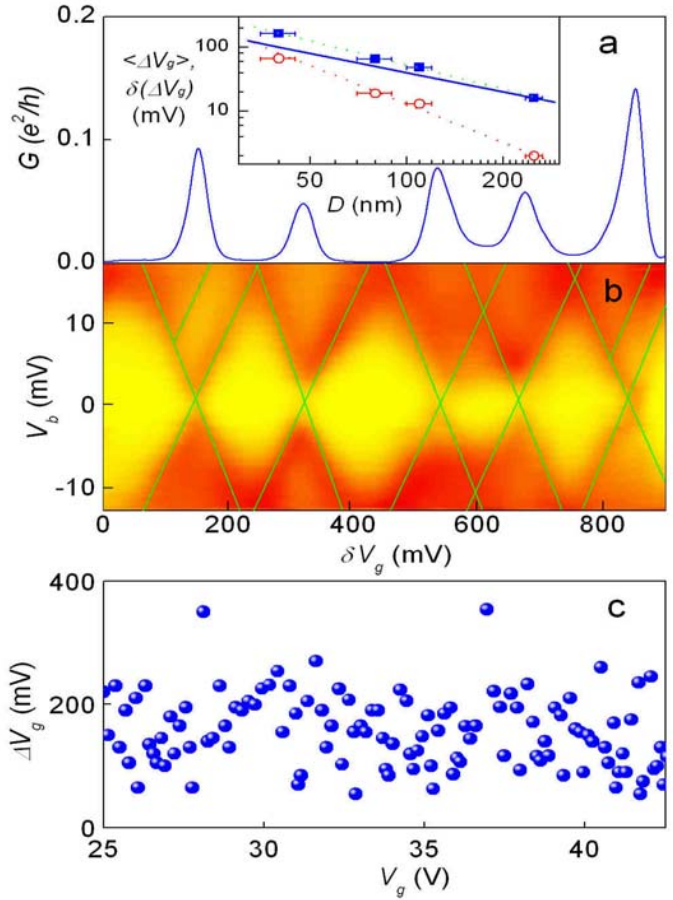

Figure 2. Effect of quantum confinement in graphene QDs. CB peaks (**a**) and Coulomb diamonds (**b**) for a 40-nm QD ($T = 4$ K). Variations in peak spacing and the size of diamonds are clearly seen. Yellow-to-red scale in (**b**) corresponds to $G_{diff}$ varying from zero to $0.4 e^2/h$. Two excited states (marked by additional lines) are feebly visible around $\delta V_g \approx 150$ and 850mV and $V_b \approx 10$mV. The smearing is caused by a rapid increase in the transparency of quantum barriers at higher $V_b$. Other examples of excited states are given in Supplementary Information. (**c**) – Separation of nearest-neighbor peaks at low $V_b$ in the same device for a large interval of $V_g$ (beyond this interval, CB became suppressed by high transparency of the barriers). Inset in (**a**): Log-log plot of the average peak spacing $<\Delta V_g>$ (solid squares) and its standard deviation $\delta(\Delta V_g)$ (open circles) as a function of $D$. Linear ($<\Delta V_g> \propto 1/D$, solid line) and quadratic ($\delta(\Delta V_g) \propto 1/D^2$, dashed) dependences are plotted as guides to the eye. The dotted curve is the best fit for the average peak spacing: $<\Delta V_g> \propto 1/D^\alpha$ yielding $\alpha \approx 1.25$.

For four devices with different $D$, we have carried out statistical analysis of their peak spacing (Fig. 3). As QDs become smaller, the average distance $<\Delta V_g>$ between CB peaks gradually increases (Fig. 2a; inset). General expectations suggest that $<\Delta V_g>$ should be proportional to $1/D$, being determined by two contributions to the QD capacitance: geometrical and quantum [11]. According to the above formula, the geometrical capacitance is $\propto D$. The quantum capacitance is given by the confinement energy and, in the first approximation, also expected to vary as $D$. Indeed, it has been shown [15] that energy levels $E_{nl}$ of Dirac fermions inside a disk of diameter $D$ are described by $J_l(E_{nl}D/2\hbar v_F) = J_{l+1}(E_{nl}D/2\hbar v_F)$ where $n$, $l$ are the main and orbital quantum numbers, respectively, and $J_l(x)$ the Bessel functions. This equation yields a typical level splitting $<\delta E> \propto 1/D$ [15], in qualitative agreement with the behavior of $<\Delta V_g>$ in Fig. 2a.

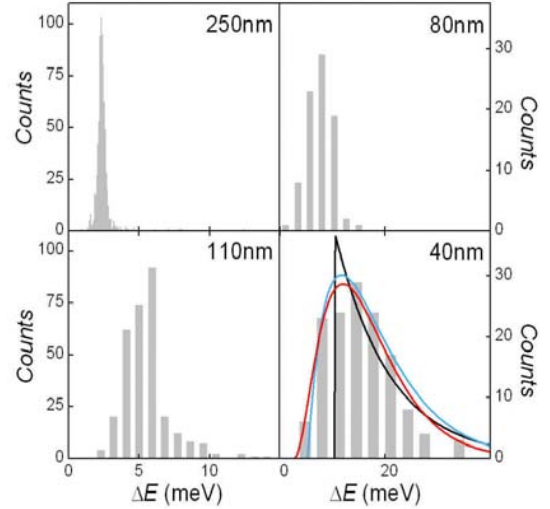

Figure 3. Level statistics. Histograms of the nearest-neighbor peak spacing $\Delta V_g \propto \Delta E$ in QDs of different $D$. The proportionality coefficient was found by measuring a number of Coulomb diamonds for each of the devices. The level statistics becomes increasingly non-Poissonian for smaller QDs. This is illustrated for the smallest device where the red, black and blue curves are the best fits for the Gaussian unitary, Poisson and Gaussian orthogonal ensembles, respectively. Note that there are no states at low $\Delta E$ because the statistical distributions are shifted from the origin due to charging energy $E_c$.

However, further analysis reveals that the above simple picture starts to break down for $D$ ~100nm. One can see from Fig. 3 that the shape of the spectral distribution notably changes: for small QDs, the histograms become increasingly broader, in comparison with their average positions. Also, $<\Delta V_g>$ changes somewhat quicker than $1/D$ (Fig. 2a; inset). We have calculated statistical deviations $\delta(\Delta V_g)$ from the average $<\Delta V_g>$ and found that $\delta(\Delta V_g)$ grows approximately as $1/D^2$ with decreasing $D$ (Fig. 2a). For example, for $D \approx 40$nm, average fluctuations in the peak spacing become as large as $<\Delta V_g>$ itself, which essentially means random positions of CB peaks. The observed behavior contradicts the one expected for Dirac fermions confined inside an ideal disk [15].

To explain this, we point out that any other confinement (except for the circular one) is predicted to result in quantum chaos. This tendency to chaos is exclusive to Dirac billiards in which even the simplest square geometry leads to chaotic trajectories [15]. In fact, the experimentally observed level statistics agrees well with the one predicted for chaotic Dirac billiards. In this case, the quantum capacitance is no longer $\propto D$ because $\delta E$ changes as $\propto 1/D^2$ [10] reflecting the fact that the level degeneracy at large $n$ and $l$ is lifted, and the number of states around a given energy is proportional to the dot area $\propto D^2$. This effect is often referred to as the level repulsion, a universal signature of quantum chaos (for both Schrödinger and Dirac billiards). The observed random spacing of CB peaks, random height of Coulomb diamonds, changes in $<\Delta V_g>$ quicker than $1/D$ and, especially, the pronounced broadening of the spectral distribution all indicate that chaos becomes a dominant factor for small QDs.

To corroborate this further, Fig. 3 shows that the observed level spacing is well described by Gaussian distribution $(32/\pi^2)\delta E^2\exp(-4\delta E/\pi)$ (characteristic of chaotic neutrino billiards) rather than the Poisson statistics $\exp(-\delta E)$ expected for integrable geometries [15]. We have also tried to distinguish between unitary and orthogonal ensembles and found that the Gaussian unitary distribution fitted our data somewhat better (Fig. 3). This agrees with theory expectations because zigzag or random edges break down the sublattice symmetry leading to the unitary statistics [15]. Further evidence for the level repulsion in small QDs is provided by the absence of any apparent bunching in their spectra (Fig. 2c). Indeed, despite considerable effort we did not succeed in finding repetitive quartets or pairs of CB peaks, which in principle could be expected due to spin and/or valley degeneracy. The latter degeneracy is lifted by scattering at QD boundaries that partially mix valleys [18], whereas the spin degeneracy may be removed by scattering on localized spins due to broken carbon bonds [5].

For even smaller devices ($D$ <30nm), the experimental behavior is completely dominated by quantum confinement. They exhibit insulating regions in $V_g$ sometimes as large as several V, and their stability diagrams yield the level spacing exceeding ~50 meV (Figs. 4a,b). However, because even the state-of-the-art lithography does not allow one to control features <10nm in size, the experimental behavior varies widely, from being characteristic of either an individual QD or two QDs in series or an individual QPC [SI]. It is also impossible to relate the observations with the exact geometry as both scanning electron and atomic force microscopy fail in visualizing the one-atom-thick elements of several nm in size and often covered by PMMA or its residue. Nevertheless, we can still use $\delta E$ to estimate the spatial scale involved. Basic arguments valid at a microscopic scale require $a/D \approx \delta E/t$ (where $a$ is the interatomic distance, $t \approx$ 3eV the hopping energy), which again yields $\delta E \approx \alpha/D$ with $\alpha \approx$ 0.5eV nm, in agreement with the literature values [3-8]. For example, for the QD shown in Fig. 4 with $\Delta E \approx$ 40meV, we find $D \sim$ 15nm.

Finally, we used our smallest devices (QDs and QPCs [SI]) to increase $\delta E$ by further decreasing their size. By using plasma etching, we could gradually narrow them by several nm from each side (in the case of QDs, one of the constrictions is then likely to dominate the whole conductance). Some of the devices became over-etched and stopped conducting but in other cases we succeeded to narrow them down to a few nm so that they exhibited the transistor action even at room $T$. Figure 4c illustrates this behavior. The device appears completely insulating with no measurable conductance ($G$ <$10^{-10}$S) over an extended range of $V_g$ (>30V) (OFF state) but then it suddenly switches ON exhibiting rather high $G \approx 10^{-3}$ $e^2/h$. At large biases, we have observed the conductance onset shifting with $V_b$ [SI], which allows an estimate for $\Delta E$ as ≈0.5eV. Importantly, this value agrees with the $T$ dependence measured near the onset of the ON state [SI], which shows that we do not deal with several QDs in series (as it was argued to be the case for nanoribbons [28]). With no possibility to control the exact geometry for the nm sizes, we cannot be certain about the origin of the observed switching. Also, the exact boundary arrangements (armchair vs zigzag vs random edge and the termination of dangling bonds) can be important at this scale [5-9]. Nevertheless, $\delta E$ ~0.5eV again allows us to estimate the spatial scale involved in the confinement as only ~1nm.

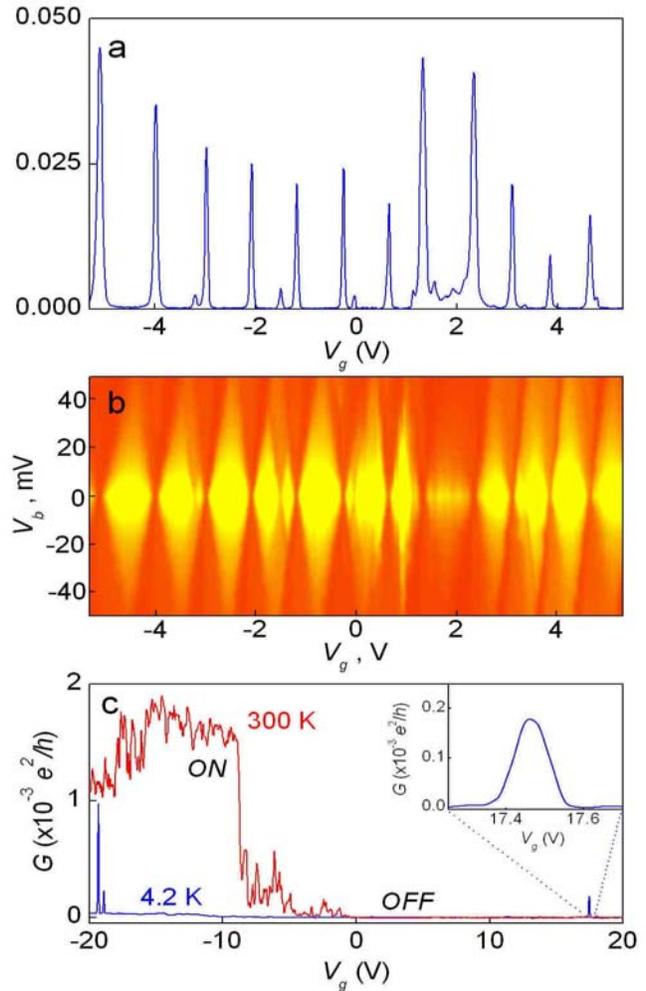

Figure 4. Electron transport through nm-scale graphene devices. Coulomb blockade peaks (**a**) and diamonds (**b**) for a QD with an estimated size ~15 nm. **c**, Electron transport through a controllably-etched device with a minimal width of only ≈1nm as estimated from its $\Delta E$. Its conductance can be completely pinched-off even at room $T$. Fluctuations in the ON state at room $T$ are time dependent (excess noise). At low $T$, the ON state exhibits much lower $G$ and the noise disappears, giving way to transmission resonances as magnified in the inset.

In conclusion, graphene QDs are an interesting and versatile experimental system allowing a range of operational regimes from conventional single-electron detectors to "neutrino billiards", in which size effects are exceptionally strong and chaos develops at relatively large diameters. Despite the fact that modern technology does not yet allow control of device's geometry on the true nm scale, we have demonstrated that graphene – unlike any other bulk material – remains stable, robust and highly conductive at the essentially molecular scale (a few benzene rings), which improves prospects of graphene-based electronics if or when the nm-scale processing techniques are developed.

Acknowledgements – The research was supported by EPSRC (UK) and the Royal Society. We are grateful to K. Ensslin, L. Eaves, M. Berry, L. Vandersypen, A. Morpurgo, A. Castro Neto, F. Guinea and M. Fromhold for helpful discussions.

## Supplementary Information

**Quantum dot designs**

The SEM micrograph in Fig. 1 of the main text shows the basic QD design but is also intended to illustrate the high resolution of our electron-beam lithography that allows features as small as 10 nm to be made reliably. The shown arrangements of side electrodes would be standard for most semiconductor-quantum-dot experiments [11,14,26]. However, we have found this design to be less efficient for our studies that required sweeps over hundreds of energy levels. Indeed, if we placed side gates in the immediate proximity of CIs (like shown in Fig. 1), this limited the number of levels that could be probed inside our smallest QDs usually down to ~10 ($\Delta V_g$ were up to 10V). We attribute the reduced influence of the back gate to additional strong screening by side electrodes. The adjacent side gates were also not efficient enough to scan over many quantum states because of the electrical breakdown or leakage along the device surface. To probe hundreds of energy levels needed for the analysis presented in the main text, we often employed the geometry shown in Fig. S1, in which side gates were located ~1 μm away from the CI. The back-gate voltage was swept typically between ±60V (limited by the onset of a leakage current through the gate dielectric), and a distant side gate was used for scans over small energy intervals only.

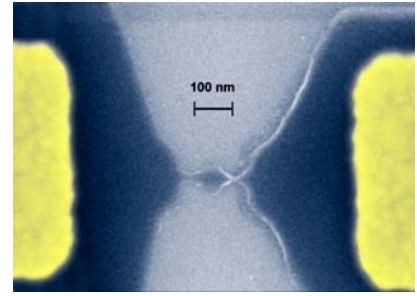

Figure S1. The scanning electron micrograph (in false color) illustrates the most typical design of our quantum dot devices. In this case ($D \approx$80nm), side-gate electrodes are remote and outside the frame. Their removal from the immediate vicinity of the CI significantly improves its coupling to the back gate allowing scans over hundreds rather than dozens of CB peaks for devices smaller than ~100nm. Dark areas in the SEM micrograph are the PMMA mask that protected graphene during etching so that it is left only underneath the mask. Yellowish regions are Au/Ti contacts.

**Graphene-based quantum point contacts**

The graphene constrictions that define quantum dots in our experiments were normally designed to have equal length and width $w$. We have avoided long constrictions as they often result in the development of spurious quantum dots within individual constrictions [28,S1]. Wide constrictions ($w$ >40nm) have too high conductance and do not allow Coulomb blockade, whereas narrow ones ($w$ <10nm) exhibit large energy gaps with no measurable conductance (see below) and, therefore, no possibility to probe QDs. From experience, we have found $w \approx$20nm to be most suitable for making quantum barriers with $G$ less but not much less than $e^2/h$, which is optimal for CB measurements. For a reference, Figure S2 shows the typical behavior exhibited by such individual constrictions that can also be referred to as quantum point contacts (QPC). The overall shape of their $G(V_g)$-response resembles $\sigma(V_g)$ for bulk graphene with the neutrality point (NP) shifted by chemical doping (to +50V in Fig. S2). It is obvious that the behavior of QPCs (with their high $G$ over a large range of back-gate voltages) limits the operation of our quantum dots to regions of less than ~20V around the $G$ minima, where QPCs remain sufficiently low conductive (compare Figs. 1b and S2).

Individual 20-nm constrictions exhibit mesoscopic (interference) fluctuations as a function of $V_g$ (see Fig. S2) but they are normally smooth on a typical scale of CB oscillations. Also, with increasing bias $V_b$, $G(V_g)$-curves become smoother with the minimum in $G(V_g)$ becoming increasingly less pronounced, which replicates the behavior observed with increasing $T$. Importantly, no Coulomb diamonds have been observed for such relatively wide QPCs above 4 K (see further), in contrast to the behavior reported for graphene nanoribbons [3,4,S1] (i.e. in our case, fluctuations decrease in amplitude with increasing $V_b$ rather than broaden). This shows that individual QPCs (together with the source and drain regions) are only responsible for smooth variations in the height of CB peaks (see Fig. 1) and cannot possibly cause random peak positions, which would require several QDs in series of approximately the same size, as in the case of stochastic CB [27] (see below).

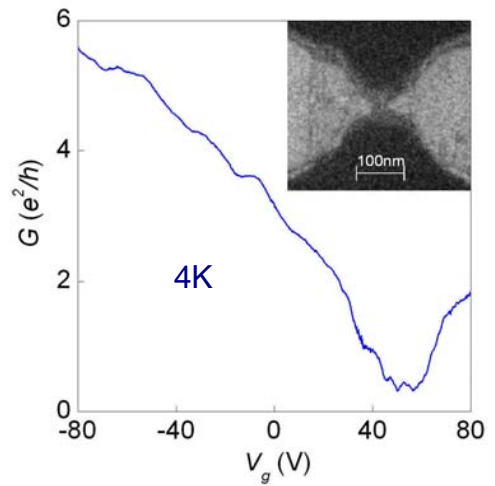

Figure S2. Conductance of an individual 20-nm graphene constriction, similar to those used in our quantum dots. Making them short and narrow has ensured that no spurious quantum dots were present inside the constrictions.

**Nanometer-sized point contacts**
We have also studied QPCs of smaller sizes (design $w$ of ≈10 nm). The inset in Figure S3 shows a micrograph for one of such nm-scale devices. With changing gate voltage, they usually exhibit two well-defined regimes: complete pinch-off with conductance below our detection limit of $G$ <$10^{-10}$ S and a strongly fluctuating finite $G$. Both regimes persist over extremely large intervals of $V_g$. This obviously makes so narrow QPCs not suitable for the QD design. Figure S3 shows that fluctuations in the conducting regime are suppressed at finite biases. Again, no diamonds are observed in this regime but instead the amplitude of fluctuations decreases with increasing $V_b$, which is typical for interference phenomena. In contrast, the pinch-off region gradually becomes narrower with increasing $V_b$ so that only one but huge Coulomb diamond could be seen in this regime. For the case of Figure S3, the size of the diamond yields a confinement gap $\delta E$ of ≈150meV. Note that, after oxygen plasma etching, our QPCs become somewhat narrower (by a few nm) than their designed width of 10nm. We have also measured the $T$ dependence of conductivity in the pinch-off regime (inset in Fig. S3) and found an activation gap $E_A$ ~80meV. The latter is approximately twice smaller than the gap found from the size of the diamond, which can be explained by likely impurity states within the confinement gap [S2].

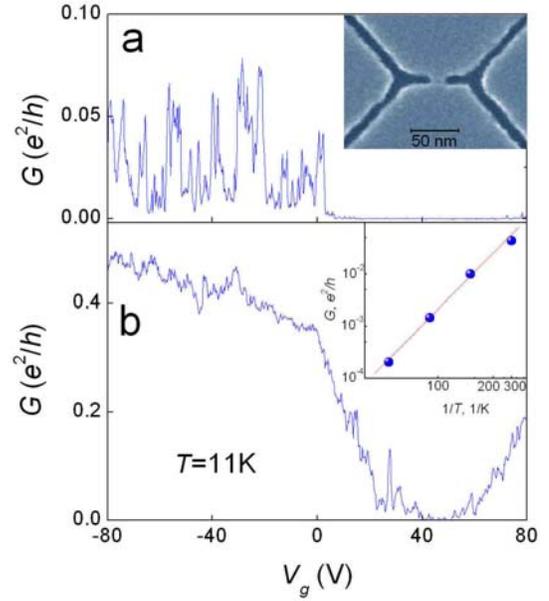

Figure S3. Electron transport through 10-nm graphene constrictions. (**a,b**) – conductance $G$ as a function of back-gate voltage at zero and 100 mV bias, respectively. Mesoscopic fluctuations in the conducting regime become suppressed, whereas the pinch-off region becomes narrower, indicating a large confinement gap. Inserts: (**a**) – SEM micrograph of one of our narrow QPCs (before plasma etching); and (**b**) – $G$ in the pinch-off region ($V_g$≈40V) can be fitted by the activation $T$ dependence $G \propto \exp(-E_A/2T)$.

**Excited states in quantum dots**
The importance of quantum confinement for graphene-based quantum dots is also witnessed through the presence of excited states on the stability diagrams [11]. In Fig. 2b such states are faint (additional lines have to be drawn to indicate them). This is a common case for our quantum dots, in which graphene constrictions rapidly become more conductive under higher applied biases, as described above. Accordingly, Coulomb diamonds' boundaries away from zero $V_b$ are smeared by high transparency of the quantum barriers, which blurs diamonds at high biases (Fig. 2) and leaves little chance for excited states to be observed. However, in some cases, QPCs fortuitously remain resistive enough even at high $V_g$ and excited states could be seen rather clearly (see Fig. S4).

**Quantum dots in series**
The high accuracy of our electron-beam lithography guaranteed that there is no accidental additional confinement that would let our devices operate as two or more quantum dots in series and lead to stochastic CB [27]. Indeed, this would require a barrier between two QDs with resistance ~$h/e^2$, which as shown above is hard to achieve without an additional constriction narrower than ~30 nm. Our high-resolution lithography and following visualization in a scanning electron microscope rule out such constrictions for devices with $D$ ≥30nm (for example, see micrographs in Figs. 1 and S3). Furthermore, electron and hole puddles always present in graphene near the neutrality point [S3] cannot be responsible for the observations of random peaks for several reasons.

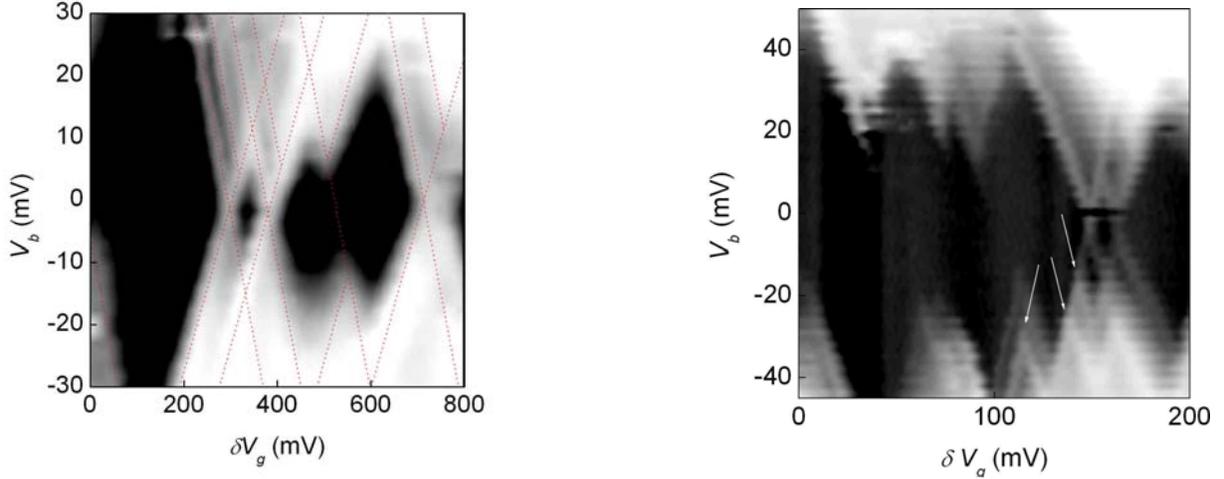

Figure S4. Examples of excited states as observed for two different quantum dots, when the barriers remained low conductive (<0.5 $e^2/h$) even at high biases. The red dashed lines (left figure) and small white arrows (on the right) are guides to the eye.

First, barriers between such puddles are expected to be rather transparent because of the Klein tunneling [1,2]. Second, the reported CB peaks extend over a large interval of gate voltages with typically more than one hundred peaks recorded whereas puddles are not deep enough to allow more that a couple of electrons inside [S3]. Still, it is instructive to compare the reported chaotic behavior of CB peaks with the one where quantum dots in series cause the stochastic CB. In the latter case, some CB peaks originating from one of the QDs completely disappear due to zero conductance through the other dots and vice versa [27]. The critical signature of two or more QDs in series is not only seemingly random positions of CB peaks (statistics would still reveal that they are not random) but also accompanying strong variations in their amplitudes (from 100% to a tiny fraction of 1%). Moreover, some peaks that are completely suppressed at low $T$ should appear at higher $T$ as the conductance of the blocking dot increases.

This CB behavior is illustrated in Fig. S5. In this case, because of a mistake in lithography, a relatively large QD appeared in series with a smaller QD ($D \approx 20$nm) inside one of the barriers (as found in SEM). One can see a seeming random pattern of CB peaks in Fig. S5. However, more careful inspection reveals that many of the peaks have a common spacing, and most of the larger gaps correspond to the double spacing, which suggests one of the peaks missing. When we increase $T$ or $V_g$, the missing peaks dutifully reveal themselves, as expected for stochastic CB [27]. This behavior of double QDs is in stark contrast to the one observed in our small individual QDs, in which – despite the absence of periodicity – CB peaks exhibit only smooth variations in their height, no additional peaks appeared at higher biases or with increasing $T$ and, probably most convincingly, the $T$ dependence of CB was in agreement with the measured size of Coulomb diamonds.

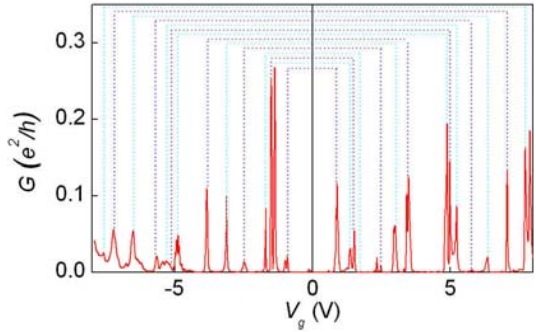

Figure S5. Stochastic Coulomb blockade. Two quantum dots in series can lead to seemingly random peaks because some of them essentially disappear whereas others have tiny amplitudes. We also note that, in rare cases, we succeeded in finding certain symmetry for electron and hole peaks. This is the case of this particular device, where the neutrality point (revealed by a clear minimum in $G$ at room $T$) is at $V_g \approx 0$V and the peaks are positioned symmetrically with respect to NP as indicated by dotted lines.

S1. B. Ozyilmaz, P. Jarillo-Herrero, D. Efetov, P. Kim. *Appl. Phys. Lett.* **91**, 192107 (2007).
S2. D. Basu, M.J. Gilbert, L.F. Register, A.H. MacDonald, S.K. Banerjee. arXiv:0712.3068.
S3. J. Martin *et al*. *Nature Phys*. doi:10.1038/nphys781 (published online Nov 2007).